\documentclass[10pt,journal]{IEEEtran}

\newcommand{\m}[1]{\ensuremath{#1}}


\usepackage{cite}
\usepackage{float}
\usepackage{amsmath,amssymb,amsfonts,amsthm}

\usepackage{tikz}
\usetikzlibrary{decorations.pathreplacing} 
\usepackage{graphicx}
\usepackage{booktabs}
\usepackage{pifont}
\usepackage{multirow}
\usepackage{siunitx}
\usepackage{makecell}

\usepackage[ruled,linesnumbered]{algorithm2e}
\usepackage{bm} 

\SetKwSty{textbf}
\SetKwInput{KwIn}{\textbf{Input}} 
\SetKwInput{KwOut}{\textbf{Output}}

\SetCommentSty{textnormal}

\usepackage{subcaption}
\usepackage{textcomp}
\usepackage{xcolor}
\usepackage{array}
\usepackage{stfloats}
\usepackage{url}
\usepackage{verbatim}
\usepackage{orcidlink}
\usepackage{hyperref}
\usepackage{tabularx}

\SetAlCapFnt{\normalsize}
\SetAlCapNameFnt{\normalsize}

\hypersetup{%
    colorlinks,
    citecolor=blue,    
    linkcolor=blue,   
    urlcolor=blue,   
    hypertexnames=false,
}

\def\BibTeX{{\rm B\kern-.05em{\sc i\kern-.025em b}\kern-.08em
    T\kern-.1667em\lower.7ex\hbox{E}\kern-.125emX}}

\begin{document}
\bstctlcite{IEEEexample:BSTcontrol}
\title{Variable-Length Finite-Rate CSI Feedback With Generative Priors}

\author{
    Yangxuan Cheng,
    Fanyang Meng, \textit{Member, IEEE},
    Jian Zou,
    Jiacheng Xie,
    Zhongqiang Zhang, \\
    Ye Wang, \textit{Member, IEEE},
    and Yongsheng Liang, \textit{Member, IEEE}
    \thanks{
        \textit{(Corresponding author: Yongsheng Liang; Fanyang Meng.)}

        Yangxuan Cheng is with Pengcheng Laboratory, Shenzhen, 518060, China and the School of Information Science and Technology, Harbin Institute of Technology, Shenzhen, 518055, China (e-mail: chengyx01@pcl.ac.cn).

        Fanyang Meng, Zhongqiang Zhang and Ye Wang are with the Pengcheng Laboratory, Shenzhen, 518060, China (e-mail: mengfy@pcl.ac.cn; zhangzhq@pcl.ac.cn; wangy02@pcl.ac.cn).

        Jian Zou is with the College of Applied Technology, Shenzhen University, Shenzhen, 518060, China (e-mail: zoujian250@gmail.com).

        Jiacheng Xie is with the School of Information Science and Technology, Harbin Institute of Technology, Shenzhen, 518055, China (e-mail: 13207303668@163.com).

        Yongsheng Liang is with the School of Information Science and Technology, Harbin Institute of Technology, Shenzhen, 518055, China and is also with Guangdong Engineering Technology Research Center of Edge Intelligence, Shenzhen Technology University, Shenzhen 518118, China (e-mail: liangys@hit.edu.cn).
    }
}

\maketitle

\begin{abstract}

This letter studies scalable finite-rate CSI feedback for FDD massive MIMO. Existing scalable neural schemes usually obtain rate flexibility by ordering, masking, quantizing, vector-quantizing, or entropy-coding learned latents, which couples the finite-bit interface to a task-specific latent codec. We propose CsiCoGen, a generative feedback mechanism that moves the finite-bit decision to codebook-constrained Gaussian innovation selection along a reverse diffusion trajectory. A synchronized pseudo-random Gaussian codebook makes each index a generative update instruction; a length-\m{L} prefix uses \m{R_L=L\log_2K} bits and yields a valid CSI estimate. The codebook is training-free and not transmitted online, while the denoiser is pretrained as a shared CSI prior. On COST2100, CsiCoGen attains indoor/outdoor NMSE of \m{-28.58}/\m{-13.96} dB at \m{792} bits and \m{-30.72}/\m{-20.37} dB at \m{1592} bits, with corresponding \m{\rho} values of \m{0.9964}/\m{0.9597} and \m{0.9967}/\m{0.9748}. Accelerated-sampling throughput and MRT spectral-efficiency results further quantify the complexity and link-level effects.
\end{abstract}


\section{Introduction}\label{sec:introduction}
\IEEEPARstart{M}{assive} multiple-input multiple-output (MIMO) remains a key enabler for 5G-Advanced and emerging 6G systems~\cite{dreifuerst2023massive,chen20235g,zhang2025covertRSMAAmBC,zou2026superimposed}. 
Its gains rely on accurate downlink channel state information (CSI) at the base station (BS). In frequency division duplex (FDD) systems, the user equipment (UE) must estimate the downlink CSI and feed it back explicitly~\cite{wen2018deep,zou2025distributed}, creating a finite-rate bottleneck as antenna dimensions and bandwidth grow. Practical CSI feedback should therefore expose actual bits or indices, support partial feedback over multiple budgets, and adjust quantization granularity without retraining the whole codec for each operating point.

Scalable CSI feedback has been pursued through shared rate-adaptive models and explicit latent-coding mechanisms. Multiple-rate CsiNet supports several feedback budgets~\cite{MRCSCsiNet2020}, while changeable-rate networks with feedback-overhead control and learned variable-rate codes handle multiple operating points within a common framework~\cite{liang2022changeable,LVRCodes2022GLOBECOM}. Variable-length schemes further adapt latent dimensionality or entropy-coded representations to the available budget~\cite{nerini2022variable,MRVLCSI2024ICASSP}. Within the latent-coding line, ordered VQ supports partial feedback~\cite{rizzello2023user}, while bit-level methods optimize latent-bit selection or quantization~\cite{DUBitLevel2023WCL,ConcreteFeedbackLayer2024TWC,zhang2023quantization}. Scalar, vector, and multi-stage quantizers provide explicit finite-alphabet interfaces~\cite{liotopoulos2025multi,shin2024vector,MSVQ2024ICC}. Nonlinear transform coding, entropy-constrained VQ, and discrete latents offer further finite-rate representations~\cite{NTC2025ICC,shin2025entropy,DiscreteLatent2024COML}. Across these schemes, the finite-bit interface remains centered on the learned latent representation and its associated rate-control or coding mechanism.

\begin{figure}[t]
    \centering
    \includegraphics[width=\columnwidth]{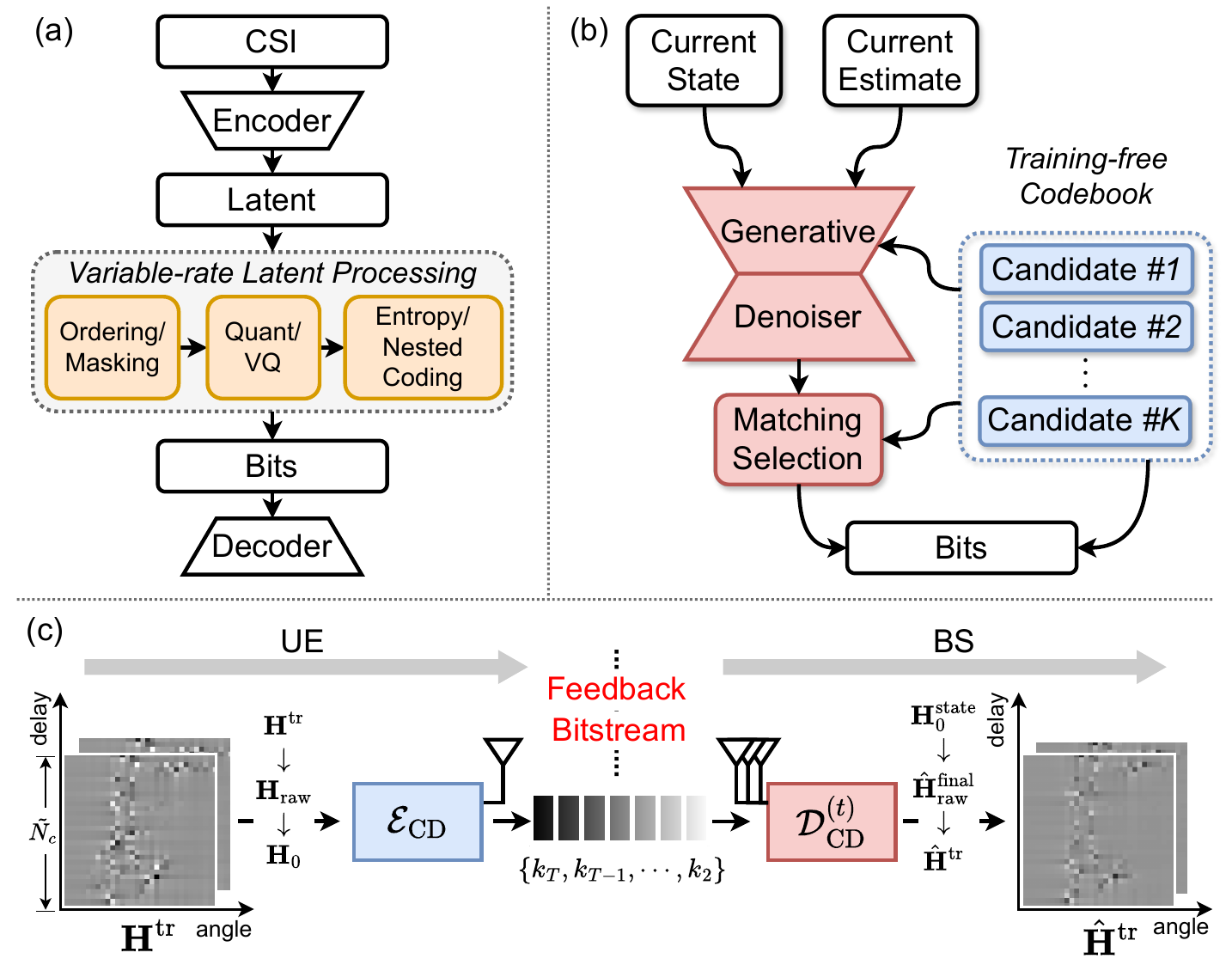}
    \caption{Mechanism-level comparison and overall workflow of CsiCoGen. (a) Existing scalable CSI feedback methods realize variable-rate operation by ordering, masking, quantizing, vector-quantizing, or entropy-coding learned latent representations. (b) CsiCoGen instead selects Gaussian innovation indices from a synchronized PRNG codebook along a shared generative recursion. (c) In the overall UE--BS workflow, the UE feeds back an index prefix and the BS progressively reconstructs CSI from the received partial sequence.}
    \label{fig:scheme}
\end{figure}

Diffusion is used here for its reverse-recursion structure, not as a generic vision-inspired enhancement. A reverse step contains a shared denoising prediction and a Gaussian innovation term. When the denoiser, schedule, initial state, and PRNG seed are synchronized at the UE and BS, the sample-dependent innovation can be restricted to a finite Gaussian codebook and represented by an index. Each received index then selects one innovation candidate and triggers one generative CSI refinement; any prefix yields a valid intermediate estimate. Without this innovation-indexing rule, the diffusion model would remain a continuous reconstructor rather than a finite-rate feedback mechanism.

In this letter, we propose CsiCoGen, a scalable finite-index CSI feedback mechanism based on codebook-constrained diffusion innovation selection. The UE selects an ordered sequence of Gaussian innovation indices, and the BS reconstructs CSI by following the same reverse recursion from a synchronized initial state. The actual feedback overhead is \m{R_L=L\log_2K}, where \m{L} is the number of received indices and \m{K} is the per-step codebook size. The PRNG Gaussian codebook is training-free and not transmitted online, while the denoiser is pretrained as the shared CSI prior. CsiCoGen therefore moves the feedback object from learned-latent quantization to reverse-diffusion innovation indexing, while related diffusion/generative CSI compression~\cite{kim2025generative} is treated as context when its conditioning assumptions or rate semantics differ from the strict finite-index benchmark used here.

\section{System Model}\label{sec:system_model}
Consider an FDD massive MIMO system with orthogonal frequency-division multiplexing (OFDM), where the BS is equipped with $N_t$ antennas and the UE has a single antenna. Let $N_f$ denote the number of OFDM subcarriers. The received signal at the UE on the $n$-th subcarrier is given by
\begin{equation}
y_n = \mathbf{h}_n^{H}\mathbf{x}_n + z_n,\qquad n=1,\dots,N_f,
\end{equation}
where $\mathbf{h}_n \in \mathbb{C}^{N_t\times 1}$ is the downlink channel vector on the $n$-th subcarrier, $\mathbf{x}_n \in \mathbb{C}^{N_t\times 1}$ is the transmitted signal, and $z_n \sim \mathcal{CN}(0,\sigma_z^2)$ is the additive noise. We assume an error-free but rate-limited uplink feedback channel.

By stacking the channel vectors of all subcarriers, the spatial-frequency CSI matrix is written as
\begin{equation}
\mathbf{H}^{\mathrm{sf}} =[\mathbf{h}_1,\mathbf{h}_2,\dots,\mathbf{h}_{N_f}]^{T} \in \mathbb{C}^{N_f\times N_t}.
\end{equation}
To exploit channel sparsity, we transform $\mathbf{H}^{\mathrm{sf}}$ into the angular-delay domain through a two-dimensional discrete Fourier transform (DFT)
\begin{equation}
\mathbf{H}^{\mathrm{ad}} = \mathbf{F}_{N_f}\mathbf{H}^{\mathrm{sf}}\mathbf{F}_{N_t}^{H},
\end{equation}
where $\mathbf{F}_{N_f}$ and $\mathbf{F}_{N_t}$ are unitary DFT matrices. Due to the limited multipath delay spread, only the first $\tilde{N}_c$ rows of $\mathbf{H}^{\mathrm{ad}}$ contain dominant channel energy ($\tilde{N}_c \ll N_f$). We retain these rows to obtain the truncated CSI matrix
\begin{equation}
\mathbf{H}^{\mathrm{tr}} = \mathbf{H}^{\mathrm{ad}}(1:\tilde{N}_c,:) \in \mathbb{C}^{\tilde{N}_c\times N_t}.
\end{equation}

For neural processing, the truncated complex-valued CSI is converted into a raw real-valued matrix by stacking its real and imaginary parts
\begin{equation}
\mathbf{H}_{\mathrm{raw}} =
\begin{bmatrix}
\Re{\mathbf{H}^{\mathrm{tr}}} \\
\Im{\mathbf{H}^{\mathrm{tr}}}
\end{bmatrix}
\in \mathbb{R}^{2\tilde{N}_c\times N_t}.
\end{equation}

In our implementation, the aforementioned DFT, truncation, and stacking steps are performed offline. The raw matrix is then globally standardized using the training-set mean $\mu_{\mathrm{train}}$ and standard deviation $\sigma_{\mathrm{train}}$ to yield $\mathbf{H}_0$, which serves as the direct explicit CSI target for our CsiCoGen compression pipeline.

Accordingly, our goal is not to learn another fixed-rate latent codec, but to define an explicit finite-index interface for a shared generative recursion. In CsiCoGen, the UE-side mapping \m{\mathcal{E}_L(\cdot)} outputs \m{L} indices from \m{\mathbf{H}_0}, and the BS-side decoder \m{\mathcal{D}_L(\cdot)} reconstructs from any received prefix under the bit budget \m{R_L=L\log_2K}.

\section{Proposed Solution}\label{sec:proposed_solution}

\subsection{Variable-Length Generative Feedback Structure}\label{sec:proposed_structure}

Fig.~\ref{fig:scheme}(c) illustrates the UE--BS workflow of CsiCoGen. The online task is to represent \m{\mathbf{H}_0} by a progressive sequence of discrete innovation indices so that the BS can decode from any prefix. Under common randomness, both ends share a synchronized Gaussian codebook \m{\mathcal{C}_t} generated by a pseudo-random number generator (PRNG).

To specify the BS-side recursive decoder, we adopt a reverse diffusion recursion with $\mathbf{H}_0$-prediction. Let the forward continuous diffusion process be defined as
\begin{equation}
    \mathbf{H}_t = \sqrt{\bar{\alpha}_t}\mathbf{H}_0 + \sqrt{1-\bar{\alpha}_t}\bm{\epsilon}, \qquad \bm{\epsilon} \sim \mathcal{N}(0,\mathbf{I}),
\end{equation}
where $\bar{\alpha}_t = \prod_{s=1}^{t}\alpha_s$ follows a predefined noise schedule. A decoder-side denoiser predicts the clean CSI from the noisy observation
\begin{equation}
    \hat{\mathbf{H}}_{0|t} = f_{\theta}(\mathbf{H}_t,t).
\label{eq:h0_prediction}
\end{equation}

Using the standard $\mathbf{H}_0$-prediction form of the diffusion reverse mean, the state update is formulated as
\begin{equation}
    \mathbf{H}_{t-1} = \bm{\mu}_t(\mathbf{H}_t,\hat{\mathbf{H}}_{0|t}) + \sigma_t\bm{\xi}_t,
\label{eq:reverse_update_cont}
\end{equation}
with the deterministic mean given by
\begin{equation}
    \bm{\mu}_t(\mathbf{H}_t,\hat{\mathbf{H}}_{0|t}) = \frac{1}{\sqrt{\alpha_t}}\left[\mathbf{H}_t + \frac{1-\alpha_t}{1-\bar{\alpha}_t}\left(\sqrt{\bar{\alpha}_t}\hat{\mathbf{H}}_{0|t}-\mathbf{H}_t\right)\right].
\label{eq:reverse_mean}
\end{equation}

The perturbation scale $\sigma_t$ adopts the posterior standard deviation $\sigma_t = \sqrt{\tilde{\beta}_t}$, where $\tilde{\beta}_t = \frac{1-\bar{\alpha}_{t-1}}{1-\bar{\alpha}_t}\beta_t$ for $t > 1$. For the final transition step ($t=1$), the update is deterministic, i.e., $\sigma_1 = 0$.

To digitize the stochastic part of the recursion, CsiCoGen uses a shared Gaussian codebook $\mathcal{C}_t$ for each reverse step,
\begin{equation}
    \mathcal{C}_t=\{\mathbf{C}_t(1),\mathbf{C}_t(2),\dots,\mathbf{C}_t(K)\}, \quad \mathbf{C}_t(k) \sim \mathcal{N}(0,\mathbf{I}).
\end{equation}

With the common PRNG seed, both ends regenerate \m{\{\mathcal{C}_t\}_{t=1}^{T_{\max}}} and \m{\mathbf{H}_{T_{\max}}} locally, so no extra runtime signaling is required beyond the feedback indices themselves. The meaningful reverse update is therefore driven by the transmitted index \m{k_t} for \m{t \in \{T_{\max},\dots,2\}},
\begin{equation}
    \mathbf{H}_{t-1} = \bm{\mu}_t(\mathbf{H}_t,\hat{\mathbf{H}}_{0|t}) + \sigma_t\mathbf{C}_t(k_t).
\label{eq:indexed_update}
\end{equation}

From a communication viewpoint, the UE selects $k_t$ through a greedy local surrogate. Specifically, at reverse step $t$, it chooses the codeword that best matches the instantaneous residual term, rather than directly optimizing the end-to-end distortion after all remaining reverse steps
\begin{equation}
    k_t = \mathop{\arg\min}\limits_{k \in \{1,\dots,K\}} \left\|\mathbf{H}_0-\hat{\mathbf{H}}_{0|t} - \sigma_t\mathbf{C}_t(k)\right\|_F^2.
\end{equation}

For isotropic Gaussian codewords, $\|\mathbf{C}_t(k)\|_F^2$ concentrates around a common value, so the above objective can be efficiently implemented by the maximum inner-product search
\begin{equation}
    k_t = \mathop{\arg\max}\limits_{k \in \{1,\dots,K\}}\left\langle \mathbf{C}_t(k), \mathbf{H}_0-\hat{\mathbf{H}}_{0|t}\right\rangle_F.
\label{eq:index_select}
\end{equation}

Since the initial state \m{\mathbf{H}_{T_{\max}}} is fully synchronized, no initial transmission is required. The final transition at \m{t=1} is deterministic and is therefore index-free. The full feedback sequence is \m{\{k_{T_{\max}},\dots,k_2\}}, but an operating point may transmit only a length-\m{L} prefix. The actual feedback overhead is
\begin{equation}
\begin{aligned}
    &0\le L\le T_{\max}-1,\qquad R_L=L\log_2K.
\end{aligned}
\label{eq:prefix_bits}
\end{equation}

Thus, \m{T_{\max}} fixes the reverse trajectory, while \m{L} fixes the transmitted prefix length; at the same \m{R_L} and \m{K}, two trajectory configurations transmit the same number of indices. After receiving \m{\{k_{T_{\max}},\dots,k_{T_{\max}-L+1}\}}, the BS runs the corresponding \m{L} updates. With real CSI dimension \m{d=2\tilde{N}_cN_t}, UE matching costs \m{O(LKd)} plus the selected denoiser evaluations, and synchronization requires only the common seed.

To make the order explicit in the retained channel dimensions, let \m{N_{\rm CSI}=\tilde N_cN_t} denote the number of retained complex angular-delay CSI coefficients and \m{d=2N_{\rm CSI}} denote the real implementation dimension. The codebook-matching step evaluates \m{K} real \m{d}-dimensional Gaussian innovation candidates, while the state update is elementwise over the same representation. After absorbing constant real/imaginary factors into big-\m{O} notation,
\begin{equation}
\begin{aligned}
C_{\rm match}
&=O(Kd)=O(K\tilde N_cN_t),\\
C_{\rm upd}
&=O(d)=O(\tilde N_cN_t),\\
C_{\rm full}
&=O\!\left(T_{\max}\big(2C_{\rm net}(\tilde N_cN_t)+K\tilde N_cN_t+\tilde N_cN_t\big)\right),\\
C_{\rm D/F}
&=O\!\left(2D C_{\rm net}(\tilde N_cN_t)+F K\tilde N_cN_t+F\tilde N_cN_t\right),\\
R_L
&=(F-1)\log_2K.
\end{aligned}
\label{eq:accelerated_complexity}
\end{equation}
Here, \m{C_{\rm net}(\tilde N_cN_t)} denotes one denoiser forward pass on the two-channel real-valued representation of the retained \m{\tilde N_c\times N_t} CSI matrix. Thus, the non-network codebook search scales linearly with both \m{K} and \m{\tilde N_cN_t}, while the state-update part scales linearly with \m{\tilde N_cN_t}. The \m{D/F} schedule keeps trajectory parameter \m{F}, corresponding to \m{F-1} index-carrying stochastic steps at the full prefix, and reduces only the number of denoiser evaluations to \m{D}; the feedback-index trajectory and bit semantics remain unchanged.

\begin{algorithm}[htb!]
\caption{Compact CsiCoGen feedback procedure}
\label{alg:CsiCoGen}
\SetAlgoLined
\KwIn{$\mathbf{H}_0$, $f_\theta$, $\{\alpha_t\}_{t=1}^{T_{\max}}$, $\mu_{\mathrm{train}}$, $\sigma_{\mathrm{train}}$, $K$, synchronized PRNG seed, received prefix length \m{L}}

\BlankLine
\textbf{UE-side progressive encoding}\;
Generate $\{\mathcal{C}_t\}_{t=1}^{T_{\max}}$ and $\mathbf{H}_{T_{\max}}$ via synchronized PRNG\;
\For{$t = T_{\max}, T_{\max}-1, \dots, 2$}{

Compute \m{\hat{\mathbf{H}}_{0|t}} and \m{\bm{\mu}_t} by \eqref{eq:h0_prediction}--\eqref{eq:reverse_mean}\;
Select \m{k_t} by \eqref{eq:index_select} and update \m{\mathbf{H}_{t-1}} by \eqref{eq:indexed_update}\;
}
For target feedback length \m{L}, the UE may terminate after selecting \m{\{k_{T_{\max}},\dots,k_{T_{\max}-L+1}\}}; the full loop is used only to generate the complete all-prefix sequence\;
\KwOut{Feedback indices $\{k_{T_{\max}},\dots,k_2\}$; a length-\m{L} prefix may be transmitted}

\BlankLine
\textbf{BS-side prefix decoding}\;
Generate $\{\mathcal{C}_t\}_{t=1}^{T_{\max}}$ and $\mathbf{H}_{T_{\max}}$ via synchronized PRNG\;
$t_L\gets T_{\max}-L$\;
If \m{L=0}, skip the received-index loop\;
\For{$t = T_{\max}, T_{\max}-1, \dots, t_L+1$}{

Use received \m{k_t} to compute \m{\hat{\mathbf{H}}_{0|t}}, \m{\bm{\mu}_t}, and \m{\mathbf{H}_{t-1}} by \eqref{eq:h0_prediction}--\eqref{eq:indexed_update}\;
}

If \m{t_L>1}, set \m{\widehat{\mathbf{H}}_{0}^{(L)}\gets f_\theta(\mathbf{H}_{t_L},t_L)}; if \m{t_L=1}, apply the deterministic final transition\;

$\hat{\mathbf{H}}_{\mathrm{raw}}^{(L)} \gets \widehat{\mathbf{H}}_{0}^{(L)} \cdot \sigma_{\mathrm{train}} + \mu_{\mathrm{train}}$\;
\KwOut{$\hat{\mathbf{H}}_{\mathrm{raw}}^{(L)}$}
\end{algorithm}

\subsection{Lightweight Generative Model Instantiation}\label{sec:proposed_model}

The proposed feedback structure only requires a decoder that follows the same \m{\mathbf{H}_0}-prediction recursion. We instantiate \m{f_\theta} as a lightweight time-conditioned residual-attention denoiser that maps \m{(\mathbf{H}_t,t)} to \m{\hat{\mathbf{H}}_{0|t}} and contains \m{800{,}162} trainable parameters; the CsiCoGen-Lite check uses a \m{155}K-parameter denoiser without changing the feedback rule.

The denoiser is trained with a weighted $\mathbf{H}_0$-prediction objective
\begin{equation}
    \mathcal{L}
    =
    \mathbb{E}_{\mathbf{H}_0,\bm{\epsilon},t}
    \!\left[
    \lambda_t
    \left\|
    f_\theta(\mathbf{H}_t,t)-\mathbf{H}_0
    \right\|_F^2
    \right],
    \label{eq:ldm_loss}
\end{equation}
where $\mathbf{H}_t=\sqrt{\bar{\alpha}_t}\mathbf{H}_0+\sqrt{1-\bar{\alpha}_t}\bm{\epsilon}$ and $\lambda_t$ emphasizes the low-noise reverse steps that dominate final reconstruction quality.

The feedback codebook is fully decoupled from denoiser training: it is generated offline by PRNG, adds no trainable parameters, and can be refreshed without modifying the denoiser. Changing \m{K} adjusts the number of innovation candidates per index, and the same online compression rule remains compatible with scene-specific denoisers.

The same denoiser is deployed at both ends, and the PRNG codebook adds no learned parameters.

\section{Simulation Results}\label{sec:simulation_results}

\begin{figure*}[!t]
    \centering
    \begin{subfigure}[t]{0.32\textwidth}
        \centering
        \includegraphics[width=\linewidth]{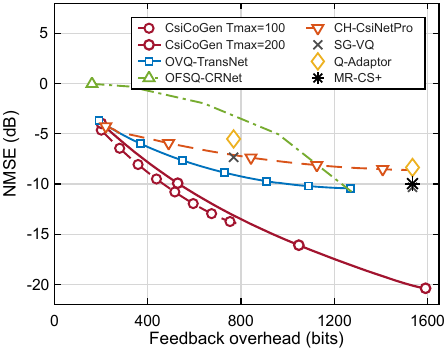}
        \vspace{-0.6em}
        \caption{Outdoor rate--NMSE benchmark.}
        \label{fig:nmse_outdoor}
    \end{subfigure}\hfill
    \begin{subfigure}[t]{0.32\textwidth}
        \centering
        \includegraphics[width=\linewidth]{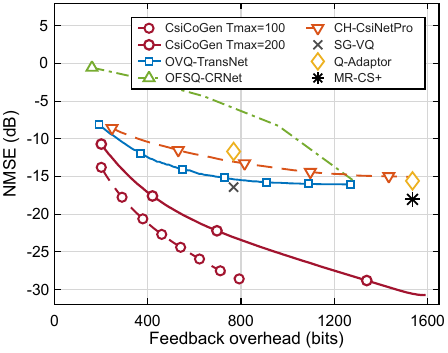}
        \vspace{-0.6em}
        \caption{Indoor rate--NMSE benchmark.}
        \label{fig:nmse_indoor}
    \end{subfigure}\hfill
    \begin{subfigure}[t]{0.32\textwidth}
        \centering
        \includegraphics[width=\linewidth]{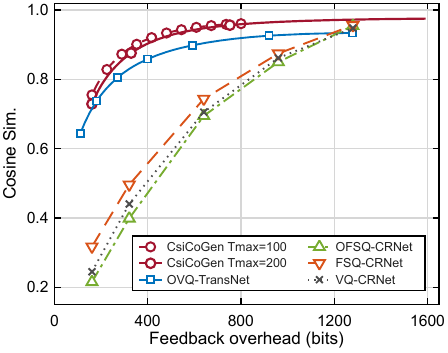}
        \vspace{-0.6em}
        \caption{Outdoor rate--\m{\rho} benchmark.}
        \label{fig:rho_outdoor}
    \end{subfigure}
    \caption{Main finite-bit COST2100 benchmark with three compact panels: (a) outdoor NMSE, (b) indoor NMSE, and (c) outdoor \m{\rho}.}
    \label{fig:main_benchmark}
\end{figure*}

We evaluate CsiCoGen on COST2100 indoor/outdoor channels~\cite{liu2012cost} with \m{100000}/\m{30000}/\m{20000} training/validation/test samples, \m{N_t=32}, \m{N_f=1024}, and \m{\tilde N_c=32}. Unless otherwise specified, \m{K=256}, so each index carries \m{8} bits. The \m{T_{\max}=100} and \m{T_{\max}=200} settings have full-prefix overheads of \m{792} and \m{1592} bits, and each curve point uses \m{R_L=L\log_2K}.
At fixed \m{R_L} and \m{K}, both settings transmit the same \m{L=R_L/\log_2 K} indices but terminate at different reverse states \m{t_L=T_{\max}-L}. Fig.~\ref{fig:main_benchmark} shows better reconstruction with \m{T_{\max}=100} over the shared range up to \m{792} bits; only \m{T_{\max}=200} extends to \m{1592} bits. Its lower full-prefix distortion is therefore obtained at a larger budget than the \m{792}-bit endpoint of \m{T_{\max}=100}.

We use a filtered protocol: \m{\mathcal{P}_0} contains directly comparable no-side-information, explicit finite-bit/index COST2100 results for the main ranking, and \m{\mathcal{P}_1} denotes partially matched or fixed-bitstream context. The benchmark includes MR-CS CsiNet+~\cite{MRCSCsiNet2020}, CH-CsiNetPro~\cite{liang2022changeable}, OVQ~\cite{rizzello2023user}, OFSQ~\cite{liotopoulos2025multi}, SG-VQ~\cite{shin2024vector}, and Q-Adaptor~\cite{zhang2023quantization}.

\begin{table}[!t]
\centering
\caption{Filtered COST2100 key points. Tags are defined in the text, and NMSE/\m{\rho} is reported when available.}
\label{tab:benchmark_keypoints}
\renewcommand{\arraystretch}{1.04}
\setlength{\tabcolsep}{1.7pt}
\scriptsize
\resizebox{\columnwidth}{!}{%
\begin{tabular}{@{}lcccc@{}}
\toprule
Method & Bits & Indoor & Outdoor & Tag \\
\midrule
{CsiCoGen, $T_{\max}=100$} & 792 & \textbf{-28.58/0.9964} & \textbf{-13.96/0.9597} & $\mathcal{P}_0$ \\
OVQ-TransNet~\cite{rizzello2023user} & 800 & -15.48/-- & -9.24/-- & $\mathcal{P}_0$ \\
OFSQ-CRNet~\cite{liotopoulos2025multi} & 960 & -8.28/0.9434 & -5.04/0.8499 & $\mathcal{P}_0$ \\
CH-CsiNetPro~\cite{liang2022changeable} & $\approx$0.8k & -13.27/-- & -7.37/-- & $\mathcal{P}_1$ \\
SG-VQ~\cite{shin2024vector} & 768 & -16.41/-- & -7.34/-- & $\mathcal{P}_0$ \\
Q-Adaptor~\cite{zhang2023quantization} & 768 & -11.68/-- & -5.513/-- & $\mathcal{P}_0$ \\
\midrule
{CsiCoGen, $T_{\max}=200$} & 1592 & \textbf{-30.72/0.9967} & \textbf{-20.37/0.9748} & $\mathcal{P}_0$ \\
OVQ-TransNet~\cite{rizzello2023user} & 1280 & -16.04/0.9863 & -10.43/0.9338 & $\mathcal{P}_0$ \\
OFSQ-CRNet~\cite{liotopoulos2025multi} & 1280 & -15.48/0.9861 & -10.87/0.9546 & $\mathcal{P}_0$ \\
CH-CsiNetPro~\cite{liang2022changeable} & $\approx$1.5k & -15.03/-- & -8.61/-- & $\mathcal{P}_1$ \\
MR-CS+~\cite{MRCSCsiNet2020} & 1536 & -18.03/-- & -9.96/-- & $\mathcal{P}_1$ \\
Q-Adaptor~\cite{zhang2023quantization} & 1536 & -15.58/-- & -8.352/-- & $\mathcal{P}_0$ \\

\bottomrule
\end{tabular}}
\end{table}

\begin{table}[!t]
\centering
\caption{1592-bit complexity, outdoor accuracy, and 10 dB MRT check. All accelerated rows use trajectory parameter \m{F=200}, corresponding to \m{199} transmitted indices at the full prefix, and reduce only denoiser calls \m{D}. Encode/decode throughput in samples/s excludes offline metric evaluation; SGCS/SE uses the specified unit-norm MRT evaluator.}
\label{tab:complexity_throughput}
\renewcommand{\arraystretch}{1.04}
\setlength{\tabcolsep}{1.4pt}
\scriptsize
\resizebox{\columnwidth}{!}{%
\begin{tabular}{@{}lcccc@{}}
\toprule
Setting & $D$ & Encode/Decode Throughput & NMSE (dB) & SGCS/SE \\
\midrule
Full F200 & 200 & $5.55{\times}10^1/4.80{\times}10^1$ & -20.37 &  0.9513/3.2110 \\
D20/F200 & 20 & $5.10{\times}10^2/5.03{\times}10^2$ & -19.48 &  0.9493/3.2084 \\
D10/F200 & 10 & $9.26{\times}10^2/9.04{\times}10^2$ & -17.54 &  0.9434/3.2005 \\
Lite-D20/F200 & 20 & $1.13{\times}10^3/1.28{\times}10^3$ & -18.16  & 0.9454/3.2031 \\
Lite-D10/F200 & 10 & $2.49{\times}10^3/2.24{\times}10^3$ & -16.46  & 0.9387/3.1941 \\
\bottomrule
\end{tabular}}
\end{table}

We use the normalized mean squared error (NMSE) to evaluate reconstruction accuracy
\begin{equation}
\mathrm{NMSE}=\mathbb{E}\left\{\frac{\|\hat{\mathbf{H}}-\mathbf{H}\|_F^2}{\|\mathbf{H}\|_F^2}\right\}.
\end{equation}

and the normalized channel-direction correlation
\begin{equation}
\rho=\mathbb{E}\left\{\frac{\left|\langle \hat{\mathbf{H}},\mathbf{H}\rangle_F\right|}{\|\hat{\mathbf{H}}\|_F\|\mathbf{H}\|_F}\right\},
\end{equation}
where $\langle \mathbf{A},\mathbf{B}\rangle_F=\mathrm{Tr}(\mathbf{A}^H\mathbf{B})$.
For downstream utility, we inverse-transform reconstructions to the spatial-frequency domain and evaluate single-user OFDM MISO MRT. For sample \m{m} and subcarrier \m{n}, let \m{\mathbf{h}_{m,n}\in\mathbb{C}^{N_t\times 1}} and \m{\widehat{\mathbf{h}}_{m,n}\in\mathbb{C}^{N_t\times 1}} denote the true and reconstructed channel vectors. For each nonzero pair, define
\begin{equation}
c_{m,n}=
\frac{\left|\mathbf{h}_{m,n}^{H}\widehat{\mathbf{h}}_{m,n}\right|}
{\|\mathbf{h}_{m,n}\|_{2}\|\widehat{\mathbf{h}}_{m,n}\|_{2}},
\qquad
\widehat{\mathbf{w}}_{m,n}=
\frac{\widehat{\mathbf{h}}_{m,n}}{\|\widehat{\mathbf{h}}_{m,n}\|_{2}}.
\end{equation}
The squared generalized cosine similarity (SGCS), used in eigenvector-based CSI feedback evaluation~\cite{EVCsiNet2021WCL}, is computed as
\begin{equation}
\begin{aligned}
\mathrm{SGCS}
&=\frac{1}{MN_f}\sum_{m=1}^{M}\sum_{n=1}^{N_f}c_{m,n}^{2}\\
&=\frac{1}{MN_f}\sum_{m=1}^{M}\sum_{n=1}^{N_f}
\frac{\left|\mathbf{h}_{m,n}^{H}\widehat{\mathbf{h}}_{m,n}\right|^{2}}
{\|\mathbf{h}_{m,n}\|_{2}^{2}\|\widehat{\mathbf{h}}_{m,n}\|_{2}^{2}}.
\end{aligned}
\end{equation}
Assuming unit-power data symbols and noise variance normalized through the SNR parameter \m{\gamma}, the bandwidth-normalized spectral efficiency is evaluated as
\begin{equation}
\mathrm{SE}(\gamma)
=\frac{1}{MN_f}\sum_{m=1}^{M}\sum_{n=1}^{N_f}
\quad \log_2\left(1+\gamma\left|\mathbf{h}_{m,n}^{H}
\widehat{\mathbf{w}}_{m,n}\right|^2\right).
\end{equation}

\begin{figure}[!t]
    \centering
    \vspace{-0.5em}
    \begin{subfigure}[t]{0.49\columnwidth}
        \centering
        \includegraphics[width=1.05\linewidth]{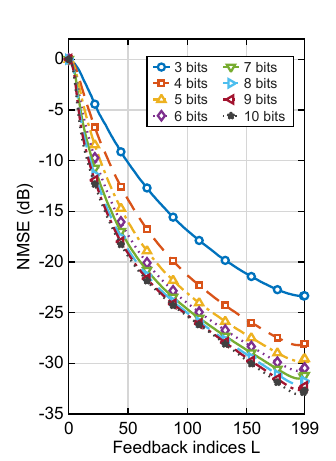}
        \vspace{-0.5em}
        \caption{Indoor.}
        \label{fig:codebook_ablation_indoor}
    \end{subfigure}
    \hfill
    \begin{subfigure}[t]{0.49\columnwidth}
        \centering
        \includegraphics[width=1.05\linewidth]{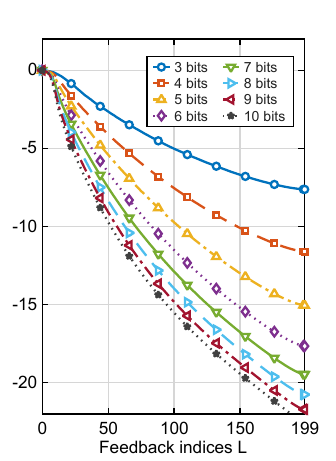}
        \vspace{-0.5em}
        \caption{Outdoor.}
        \label{fig:codebook_ablation_outdoor}
    \end{subfigure}
    \caption{Codebook-size ablation at \m{T_{\max}=200} versus prefix length \m{L}. The x-axis is \m{L}, not an equal-bit comparison; for same \m{L}, different \m{K} values use \m{R_L=L\log_2K} feedback bits.}
    \label{fig:codebook_ablation}
\end{figure}

Fig.~\ref{fig:main_benchmark} and Table~\ref{tab:benchmark_keypoints} show the filtered finite-bit COST2100 benchmark. Each additional CsiCoGen index triggers one recursive innovation update, so the curves are naturally progressive. With \m{T_{\max}=100}, the full \m{792}-bit prefix reaches \m{-28.58} dB/0.9964 indoors and \m{-13.96} dB/0.9597 outdoors in NMSE/\m{\rho}. With \m{T_{\max}=200}, the full \m{1592}-bit prefix reaches \m{-30.72} dB/0.9967 indoors and \m{-20.37} dB/0.9748 outdoors. These results are interpreted as favorable among directly comparable published no-side-information finite-bit COST2100 baselines. Large-model, side-information-assisted, average-bit, or non-COST2100 methods are kept as \m{\mathcal{P}_2} related context rather than ranked in this table. Thus, the distinction is not variable-rate capability itself, but that CsiCoGen feeds discrete innovation indices directly into the reverse diffusion recursion instead of adapting learned latent variables.

Table~\ref{tab:complexity_throughput} makes the \m{1592}-bit complexity and outdoor MRT evidence explicit. The table reports encode/decode throughput in samples/s excluding offline metric evaluation, outdoor NMSE, and 10 dB SGCS/SE under the unit-norm MRT evaluator. D20/F200 uses trajectory parameter \m{F=200}, corresponding to \m{199} full-prefix transmitted indices, and reduces only \m{D}, increasing outdoor E/D throughput from \m{5.55{\times}10^1/4.80{\times}10^1} to \m{5.10{\times}10^2/5.03{\times}10^2} samples/s with outdoor NMSE \m{-19.48} dB and SGCS/SE 0.9493/3.2084. Lite-D20/F200 and Lite-D10/F200 further reach \m{1.13{\times}10^3/1.28{\times}10^3} and \m{2.49{\times}10^3/2.24{\times}10^3} samples/s, respectively, with outdoor SGCS/SE values of 0.9454/3.2031 and 0.9387/3.1941. The outdoor case is reported as the more challenging setting; the corresponding indoor full-sampler SGCS/SE at \m{1592} bits is \m{0.9937/3.3441} and follows the same trend. Fig.~\ref{fig:codebook_ablation} uses prefix length \m{L} as the x-axis; increasing \m{K} improves per-index innovation precision, while \m{R_L=L\log_2K}, so different \m{K} values are not equal-bit comparisons at the same \m{L}.

\section{Conclusion}

This letter proposed CsiCoGen, a scalable finite-rate CSI feedback mechanism that digitizes Gaussian innovations along a reverse diffusion trajectory rather than quantizing a learned latent representation. A synchronized PRNG Gaussian codebook makes each index a generative state-update instruction, yielding prefix-controllable feedback with \m{R_L=L\log_2K}. COST2100 benchmark, throughput, and MRT spectral-efficiency results support this interpretation while quantifying the main complexity and link-level effects. The main limitation is inference cost: even with accelerated schedules, iterative diffusion reconstruction remains slower than one-shot AE, VQ, or scalar-quantization codecs. Reducing this latency through faster samplers, smaller denoisers, optimized codebook matching, and inference-runtime acceleration remains important future work.

\bibliographystyle{IEEEtran}
\bibliography{IEEEabrv,references}

@STRING{IEEE_J_VT         = "{IEEE} Trans. Veh. Technol."}

@STRING{IEEE_J_COML       = "{IEEE} Commun. Lett."}

@STRING{IEEE_J_JSAC       = "{IEEE} J. Sel. Areas Commun."}

@STRING{IEEE_J_WCOM       = "{IEEE} Trans. Wireless Commun."}

@STRING{IEEE_M_COM        = "{IEEE} Commun. Mag."}

@STRING{IEEE_M_WC         = "{IEEE} Wireless Commun."}

@STRING{IEEE_J_WCOML      = "{IEEE} Wireless Commun. Lett."}

@STRING{IEEE_C_ICC        = "Proc. {IEEE} Int. Conf. Commun. (ICC)"}

@IEEEtranBSTCTL{IEEEexample:BSTcontrol,
    CTLuse_forced_etal       = "yes",
    CTLmax_names_forced_etal = "6",
    CTLnames_show_etal       = "1"
}

@article{chen20235g,
    title={{5G-advanced} toward {6G}: Past, present, and future},
    author={Chen, Wanshi and Lin, Xingqin and Lee, Juho and Toskala, Antti and Sun, Shu and Chiasserini, Carla Fabiana and Liu, Lingjia},
    journal=IEEE_J_JSAC,
    volume={41},
    number={6},
    pages={1592--1619},
    year={2023},
    publisher={IEEE}
}

@article{dreifuerst2023massive,
    title={Massive {MIMO} in {5G}: How beamforming, codebooks, and feedback enable larger arrays},
    author={Dreifuerst, Ryan M and Heath, Robert W},
    journal=IEEE_M_COM,
    volume={61},
    number={12},
    pages={18--23},
    year={2023},
    publisher={IEEE}
}

@article{liu2012cost,
    title={The {COST} 2100 {MIMO} channel model},
    author={Liu, Lingfeng and Oestges, Claude and Poutanen, Juho and Haneda, Katsuyuki and Vainikainen, Pertti and Quitin, Fran{\c{c}}ois and Tufvesson, Fredrik and De Doncker, Philippe},
    journal=IEEE_M_WC,
    volume={19},
    number={6},
    pages={92--99},
    year={2012},
    publisher={IEEE}
}

@article{wen2018deep,
    title={Deep learning for massive {MIMO} {CSI} feedback},
    author={Wen, Chao-Kai and Shih, Wan-Ting and Jin, Shi},
    journal=IEEE_J_WCOML,
    volume={7},
    number={5},
    pages={748--751},
    year={2018},
    publisher={IEEE}
}

@article{liang2022changeable,
    title={Changeable rate and novel quantization for {CSI} feedback based on deep learning},
    author={Liang, Xin and Chang, Haoran and Li, Haozhen and Gu, Xinyu and Zhang, Lin},
    journal=IEEE_J_WCOM,
    volume={21},
    number={12},
    pages={10100--10114},
    year={2022},
    publisher={IEEE}
}

@article{MRCSCsiNet2020,
    title={Convolutional neural network-based multiple-rate compressive sensing for massive {MIMO} {CSI} feedback: Design, simulation, and analysis},
    author={Guo, Jiajia and Wen, Chao-Kai and Jin, Shi and Li, Geoffrey Ye},
    journal=IEEE_J_WCOM,
    volume={19},
    number={4},
    pages={2827--2840},
    year={2020},
    publisher={IEEE}
}

@article{rizzello2023user,
    title={User-driven adaptive {CSI} feedback with ordered vector quantization},
    author={Rizzello, Valentina and Nerini, Matteo and Joham, Michael and Clerckx, Bruno and Utschick, Wolfgang},
    journal=IEEE_J_WCOML,
    volume={12},
    number={11},
    pages={1956--1960},
    year={2023},
    publisher={IEEE}
}

@article{ConcreteFeedbackLayer2024TWC,
    title={Concrete feedback layers: Variable-length, bit-level {CSI} feedback optimization for {FDD} wireless communication systems},
    author={Ji, Dong Jin and Chung, Byung Chang},
    journal=IEEE_J_WCOM,
    volume={23},
    number={10},
    pages={15353--15366},
    year={2024},
    publisher={IEEE}
}

@article{zhang2023quantization,
    title={Quantization adaptor for bit-level deep learning-based massive {MIMO} {CSI} feedback},
    author={Zhang, Xudong and Lu, Zhilin and Zeng, Rui and Wang, Jintao},
    journal=IEEE_J_VT,
    volume={73},
    number={4},
    pages={5443--5453},
    month=apr,
    year={2024},
    doi={10.1109/TVT.2023.3333358},
    publisher={IEEE}
}

@article{shin2024vector,
    title={Vector quantization for deep-learning-based {CSI} feedback in massive {MIMO} systems},
    author={Shin, Junyong and Kang, Yujin and Jeon, Yo-Seb},
    journal=IEEE_J_WCOML,
    volume={13},
    number={9},
    pages={2382--2386},
    year={2024},
    publisher={IEEE}
}

@article{liotopoulos2025multi,
    title={Multi-Length {CSI} Feedback With Ordered Finite Scalar Quantization},
    author={Liotopoulos, Kosmas and Mitsiou, Nikos A and Sarigiannidis, Panagiotis G and Karagiannidis, George K},
    journal=IEEE_J_COML,
    volume={29},
    number={8},
    pages={1973--1977},
    month=aug,
    year={2025},
    doi={10.1109/LCOMM.2025.3581951},
    publisher={IEEE}
}

@inproceedings{kim2025generative,
    title={Generative diffusion model-based compression of {MIMO} {CSI}},
    author={Kim, Heasung and Lee, Taekyun and Kim, Hyeji and De Veciana, Gustavo and Arfaoui, Mohamed Amine and Koc, Asil and Pietraski, Phil and Zhang, Guodong and Kaewell, John},
    booktitle=IEEE_C_ICC,
    pages={6323--6328},
    year={2025}
}

@article{shin2025entropy,
    title={Entropy-constrained {VQ-VAE} for deep-learning-based {CSI} feedback},
    author={Shin, Junyong and Park, Jinsung and Jeon, Yo-Seb},
    journal=IEEE_J_VT,
    volume={74},
    number={6},
    pages={9870-9875},
    year={2025},
    publisher={IEEE}
}

@article{nerini2022variable,
    title={Machine learning-based {CSI} feedback with variable length in {FDD} massive {MIMO}},
    author={Nerini, Matteo and Rizzello, Valentina and Joham, Michael and Utschick, Wolfgang and Clerckx, Bruno},
    journal=IEEE_J_WCOM,
    volume={22},
    number={5},
    pages={2886--2900},
    year={2023},
    publisher={IEEE}
}

@article{zou2025distributed,
  title={Distributed learning-based channel estimation and feedback for {RIS}-aided wireless communications},
  author={Zou, Jian and Mao, Qingyu and Xiao, Jian and Liu, Shuai and Liang, Yongsheng},
  journal=IEEE_J_WCOML,
  volume={14},
  number={2},
  pages={460--464},
  year={2025},
  publisher={IEEE}
}

@inproceedings{LVRCodes2022GLOBECOM,
  author    = {Heasung Kim and Hyeji Kim and Gustavo de Veciana},
  title     = {Learning Variable-Rate Codes for {CSI} Feedback},
  booktitle = {Proc. {IEEE} Global Commun. Conf. (GLOBECOM)},
  restoreconfaddress = {Rio de Janeiro, Brazil},
  pages     = {1435--1441},
  restoreconfmonth   = dec,
  year      = {2022},
  doi       = {10.1109/GLOBECOM48099.2022.10000622}
}

@article{DUBitLevel2023WCL,
  author  = {Zheng Cao and Jiajia Guo and Chao-Kai Wen and Shi Jin},
  title   = {Deep-Unfolding-Based Bit-Level {CSI} Feedback in Massive {MIMO} Systems},
  journal = IEEE_J_WCOML,
  volume  = {12},
  number  = {2},
  pages   = {371--375},
  month   = feb,
  year    = {2023},
  doi     = {10.1109/LWC.2022.3227315}
}

@article{DiscreteLatent2024COML,
  author  = {Xinran Sun and Zhengming Zhang and Chunguo Li and Yongming Huang and Luxi Yang},
  title   = {An Effective Network With Discrete Latent Representation Designed for Massive {MIMO} {CSI} Feedback},
  journal = IEEE_J_COML,
  volume  = {28},
  number  = {11},
  pages   = {2648--2652},
  month   = nov,
  year    = {2024},
  doi     = {10.1109/LCOMM.2024.3462977}
}

@inproceedings{MSVQ2024ICC,
  author    = {Junho Lee and Jaein Kim and Yoojin Choi},
  title     = {Learning Variable-Rate {CSI} Compression With Multi-Stage Vector Quantization},
  booktitle = IEEE_C_ICC,
  restoreconfaddress = {Denver, CO, USA},
  pages     = {5123--5128},
  restoreconfmonth   = jun,
  year      = {2024},
  doi       = {10.1109/ICC51166.2024.10622937}
}

@inproceedings{MRVLCSI2024ICASSP,
  author    = {Bumsu Park and Heedong Do and Namyoon Lee},
  title     = {Multi-Rate Variable-Length {CSI} Compression for {FDD} Massive {MIMO}},
  booktitle = {Proc. {IEEE} Int. Conf. Acoust., Speech Signal Process. (ICASSP)},
  restoreconfaddress = {Seoul, Republic of Korea},
  pages     = {7715--7719},
  restoreconfmonth   = apr,
  year      = {2024},
  doi       = {10.1109/ICASSP48485.2024.10448212}
}

@inproceedings{NTC2025ICC,
  author    = {Bumsu Park and Heedong Do and Namyoon Lee},
  title     = {Transformer-Based Nonlinear Transform Coding for Multi-Rate {CSI} Compression in {MIMO}-{OFDM} Systems},
  booktitle = IEEE_C_ICC,
  restoreconfaddress = {Montreal, QC, Canada},
  pages     = {2327--2333},
  restoreconfmonth   = jun,
  year      = {2025},
  doi       = {10.1109/ICC52391.2025.11161672}
}

@article{zou2026superimposed,
  author  = {Jian Zou and Jian Xiao and Wenwu Xie and Fanyang Meng and Renhai Feng and Liang Yang and Yongsheng Liang},
  title   = {Superimposed Pilot-Based Adaptive Semantic Communications for Wireless Image Transmission},
  journal = IEEE_J_WCOM,
  volume  = {25},
  pages   = {17576--17590},
  year    = {2026},
  doi     = {10.1109/TWC.2026.3694102}
}

@article{EVCsiNet2021WCL,
  author  = {Wendong Liu and Wenqiang Tian and Han Xiao and Shi Jin and Xiaofeng Liu and Jia Shen},
  title   = {{EVCsiNet}: Eigenvector-Based {CSI} Feedback Under {3GPP} Link-Level Channels},
  journal = IEEE_J_WCOML,
  volume  = {10},
  number  = {12},
  pages   = {2688--2692},
  year    = {2021},
  doi     = {10.1109/LWC.2021.3112747}
}

@article{zhang2025covertRSMAAmBC,
  author  = {Zhuo Zhang and Liang Yang and Hongjiang Lei and Xingwang Li and Dusit Niyato},
  title   = {Covert Communication in {RSMA}-Assisted Ambient Backscatter Communication Systems},
  journal = IEEE_J_WCOM,
  year    = {2025},
  volume  = {24},
  number  = {8},
  pages   = {6566--6579},
  month   = aug,
  doi     = {10.1109/TWC.2025.3554667}
}

\end{document}